\documentclass[manuscript,screen]{acmart}
\usepackage{xcolor}
\usepackage{amsmath}
\usepackage{graphicx}
\usepackage{capt-of}
\usepackage{booktabs}
\usepackage{varwidth}

\DeclareMathOperator*{\argmin}{arg\,min}

\AtBeginDocument{%
  \providecommand\BibTeX{{%
    \normalfont B\kern-0.5em{\scshape i\kern-0.25em b}\kern-0.8em\TeX}}}

\copyrightyear{2021}
\acmYear{2021}
\setcopyright{rightsretained}
\acmConference[RecSys '21]{Fifteenth ACM Conference on Recommender Systems}{September 27-October 1, 2021}{Amsterdam, Netherlands}
\acmBooktitle{Fifteenth ACM Conference on Recommender Systems (RecSys '21), September 27-October 1, 2021, Amsterdam, Netherlands}\acmDOI{10.1145/3460231.3478852}
\acmISBN{978-1-4503-8458-2/21/09}

\begin{document}

\title{An Analysis Of Entire Space Multi-Task Models For Post-Click Conversion Prediction}

\author{Conor O'Brien}
\authornote{Both authors contributed equally to this research.}
\email{conoro@twitter.com}
\affiliation{
  \institution{Twitter}
  \country{UK}
}

\author{Kin Sum Liu}
\authornotemark[1]
\email{kliu@twitter.com}
\affiliation{
  \institution{Twitter}
  \country{USA}
}

\author{James Neufeld}
\email{jneufeld@twitter.com}
\affiliation{
  \institution{Twitter}
  \country{USA}
}

\author{Rafael Barreto}
\email{rbarreto@twitter.com}
\affiliation{
  \institution{Twitter}
  \country{USA}
}

\author{Jonathan J Hunt}
\email{jjh@twitter.com}

\affiliation{
  \institution{Twitter}
  \country{UK}
}

\renewcommand{\shortauthors}{C. O'Brien and K.S. Liu, et al.}
\renewcommand{\shorttitle}{Analysis of Entire Space Multi-task Models}

\begin{abstract}
Industrial recommender systems are frequently tasked with approximating probabilities for multiple, often closely related, user actions. For example, predicting if a user will click on an advertisement and if they will then purchase the advertised product.
The conceptual similarity between these tasks has promoted the use of multi-task learning: a class of algorithms that aim to bring positive inductive transfer from related tasks. Here, we empirically evaluate multi-task learning approaches with neural networks for an online advertising task. Specifically, we consider approximating the probability of post-click conversion events (installs) (CVR) for mobile app advertising on a large-scale advertising platform, using the related click events (CTR) as an auxiliary task. We use an ablation approach to systematically study recent approaches that incorporate both multitask learning and ``entire space modeling'' which train the CVR on all logged examples rather than learning a conditional likelihood of conversion given clicked. Based on these results we show that several different approaches result in similar levels of positive transfer from the data-abundant CTR task to the CVR task and offer some insight into how the multi-task design choices address the two primary problems affecting the CVR task: data sparsity and data bias. Our findings add to the growing body of evidence suggesting that standard multi-task learning is a sensible approach to modelling related events in real-world large-scale applications and suggest the specific multitask approach can be guided by ease of implementation in an existing system.
\end{abstract}

\ccsdesc[500]{Computing methodologies~Multi-task learning}
\ccsdesc[300]{Computing methodologies~Neural networks}
\ccsdesc[100]{Computing methodologies~Batch learning}

\keywords{Display advertising}

\maketitle

\section{Introduction}

Many industrial recommender system applications, particularly in the online advertising space, have tasks whose representations or labels are meaningfully related to other tasks. Sometimes this manifests as a strict causal relationship, such as a purchase event conditional on an add-to-cart action. Other times the tasks are merely correlated, for example, replying to and ``favoriting'' a social media post.

The existence of these similar events naturally raises two questions: (1) how should the objectives be modelled and (2) what is the best way to elicit positive transfer between tasks. The answers to these questions are, of course, dependent on the problem at hand. In this paper, we restrict our focus to a specific, but important, real-world scenario: predicting post-click conversion rates (CVR) for the purpose of online advertising on Twitter.

\begin{figure}[h!]
    \centering
    \includegraphics[width=0.45\textwidth]{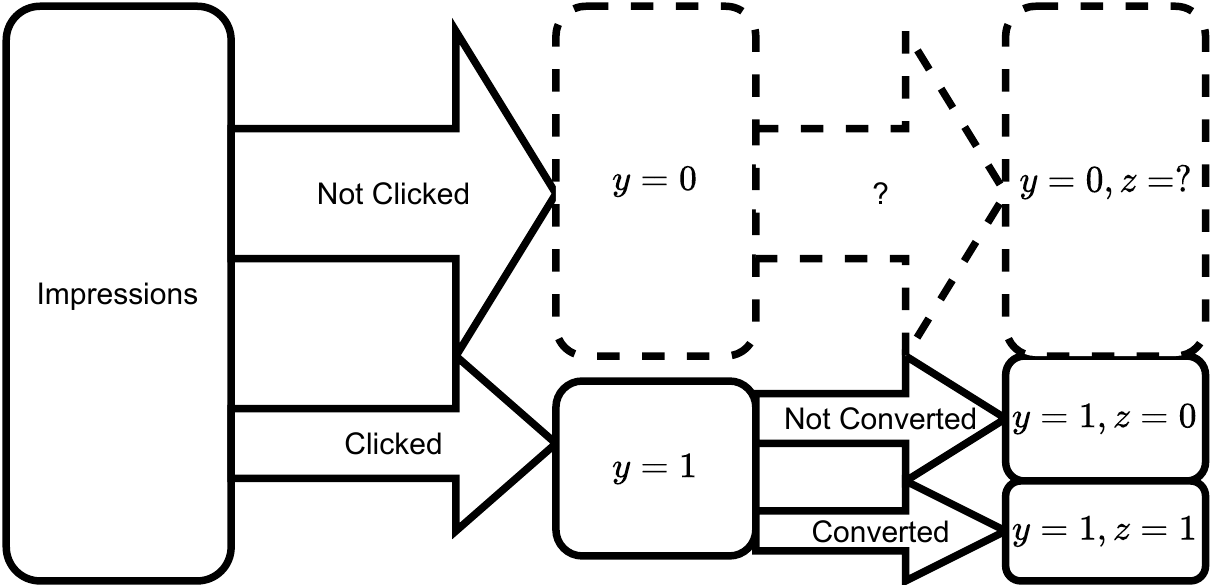}
    \caption{Event flow from impressions to clicks ($y$) to conversions ($z$). Impressions with ad-click engagements, $y=1$, are much sparser compared to those without ad-clicks, $y=0$. The dashed lines indicate conditional likelihoods that cannot be observed in practice: that is, missing counterfactuals that cause data bias. Conversion models trained only on clicked data, predicting $p(z|y=1, x)$, are affected by both the sparsity and bias when used for conversion inference on the space of all impressions.}
    \label{fig:event_flow}
\end{figure}

In the most straightforward setup, ad click-through rate (CTR) and CVR prediction are treated as separate supervised learning problems with two models trained independently. Of these tasks, CVR prediction is usually more challenging for two reasons. The first reason is \textbf{data sparsity}: every impression shown to a user generates training data for a CTR model whereas only impressions which result in a click generate training data for a CVR model. The number of impressions that generate ad-click engagements are typically a small fraction, sometimes less than 1\%, so the CVR model must be trained with significantly less data. This challenge is exacerbated by the fact that exploration data is expensive to obtain as there is opportunity cost associated with each served ad impression. Put differently, serving random traffic for better exploration comes with a significant financial disincentive. The second reason is \textbf{data bias}: the CVR model needs to make predictions over all impressions, however, only impressions which resulted in a click are used as training examples. That is, for an impression which did not result in a click, we lack the counterfactual information about whether this would have resulted in a conversion had the user clicked on the ad (see Figure \ref{fig:event_flow}).

Recent work \cite{ma2018entire} introduced an approach to modeling CVR they named Entire Space Multitask Model (ESMM) which has two key ideas: (1) sharing parameters for representation learning between the CVR and CTR problem and (2) modeling the CVR unconditionally which allows training the CVR model on all impression samples (they term ``entire space'' modeling). We expand on these descriptions below.

Here we systematically investigated, through the use of ablation studies, the mechanisms behind the good performance of the ESMM model. We reproduced the findings of \cite{ma2018entire} that ESMM outperforms modeling CVR and CTR as separate models on a different, industry scale dataset. However, we also found that a similar level of performance can be obtained by approaches which incorporated only one aspect of the ESMM model. That is, models which use only parameter sharing between CVR or CTR, or only ``entire space'' training.

\subsection{Problem Formulation}

We consider the conversion prediction problem under standard supervised learning assumptions. That is, we are assumed to be presented with an ad \emph{context}, denoted $x$, that represents the attributes of the ad placement, user request, and the ad itself, drawn i.i.d. from some stationary distribution, $D$. If this ad candidate is presented and observed by the user (an "impression") then the user will elect to \emph{click} on the ad, denoted $y = 1$, with probability $p(y = 1 | x)$. Additionally, the user may also elect to \emph{convert} by installing the advertised application or purchasing the product, denoted $z = 1$, with probability $p(z = 1 | y = 1, x)$. By construction, a conversion is only possible if the user has clicked: that is, $p(z = 1 | y = 0, x) = 0$.
The goal is to produce a classification function, $f(x; \theta) \to [0, 1]$, that minimizes the expected (cross-entropy) loss for any new example drawn from the same distribution. That is, we aim to find model parameters, $\theta$, where $\argmin_\theta \mathbb{E}_D\left[L(f(x;\theta), z)\right]$, for $L(p,z) = -(z \cdot \log(p) + (1-z) \cdot \log(1-p))$.

\subsection{Related work}

Deep learning based models have been widely studied for use in multi-task and transfer learning \cite{pmlr-v27-bengio12a, DBLP:journals/corr/YosinskiCBL14, tan2018survey, devlin2018bert, reviewer_paper1, reviewer_paper2}. One common approach to transfer learning is to share neural network parameters between related tasks, until the final hidden layer of a deep network~\cite{zhang2021survey}. The general consensus from this body of work is that relatively straightforward techniques often work well in practice and can greatly reduce the amount of time or data required to learn a new task (see survey:~\cite{zhang2021survey}). However, transfer learning can be challenging to do well, and can easily result in negative transfer if done na\"ively \cite{kirkpatrick2017overcoming, tang2020progressive}.

Regarding the specific task of conversion prediction for online advertising, the expense of obtaining labeled conversion data and the inherent rarity of successful advertising-driven conversions has encouraged the development of multi-task learning approaches in numerous industrial contexts. While the business-sensitive nature of this application does dissuade publication of production systems, there are some representative examples in the literature. For instance, as early as 2014 hierarchical multi-task learning (MTL) conversion models were deployed at scale at Yahoo~\cite{Ahmed2014}, \cite{DBLP:journals/ml/PerlichDRSP14} described a multi-task feature engineering approach for online advertising.

The approach discussed in~\cite{ma2018entire} most closely relates to the work presented in this report. There the authors addressed the problems of data sparsity in the post-click conversion task through a proposed a multi-task model sharing parameters between CTR and CVR tasks. Additionally, the authors aim to address the dataset bias issue by predicting the joint probability of click and conversion -- treating the marginal CTR prediction as an auxiliary task. This work demonstrated improved prediction performance over baselines. Also, notably, \cite{wang2020delayed} consider similar approaches for this task with a specific focus towards the issue of delayed feedback; while interesting, the challenge of delayed feedback falls outside the scope of this work.

\section{Methods} \label{sec:methods}

As detailed above, the main quantity of interest for ad ranking systems is the user-ad conversion rate, $p(z=1|x)$. There are a number of ways to decompose this prediction, which result in different characteristics and may allow for different MTL approaches. For instance, choosing to ignore the decomposition of $p(z=1|x)$, into $p(z=1|y=1,x) \cdot p(y=1|x)$, leaves only a single prediction, and hence a single-task deep neural network (DNN) architecture training in the impression space.

\begin{figure}[b]
    \centering
    \includegraphics[width=0.8\textwidth]{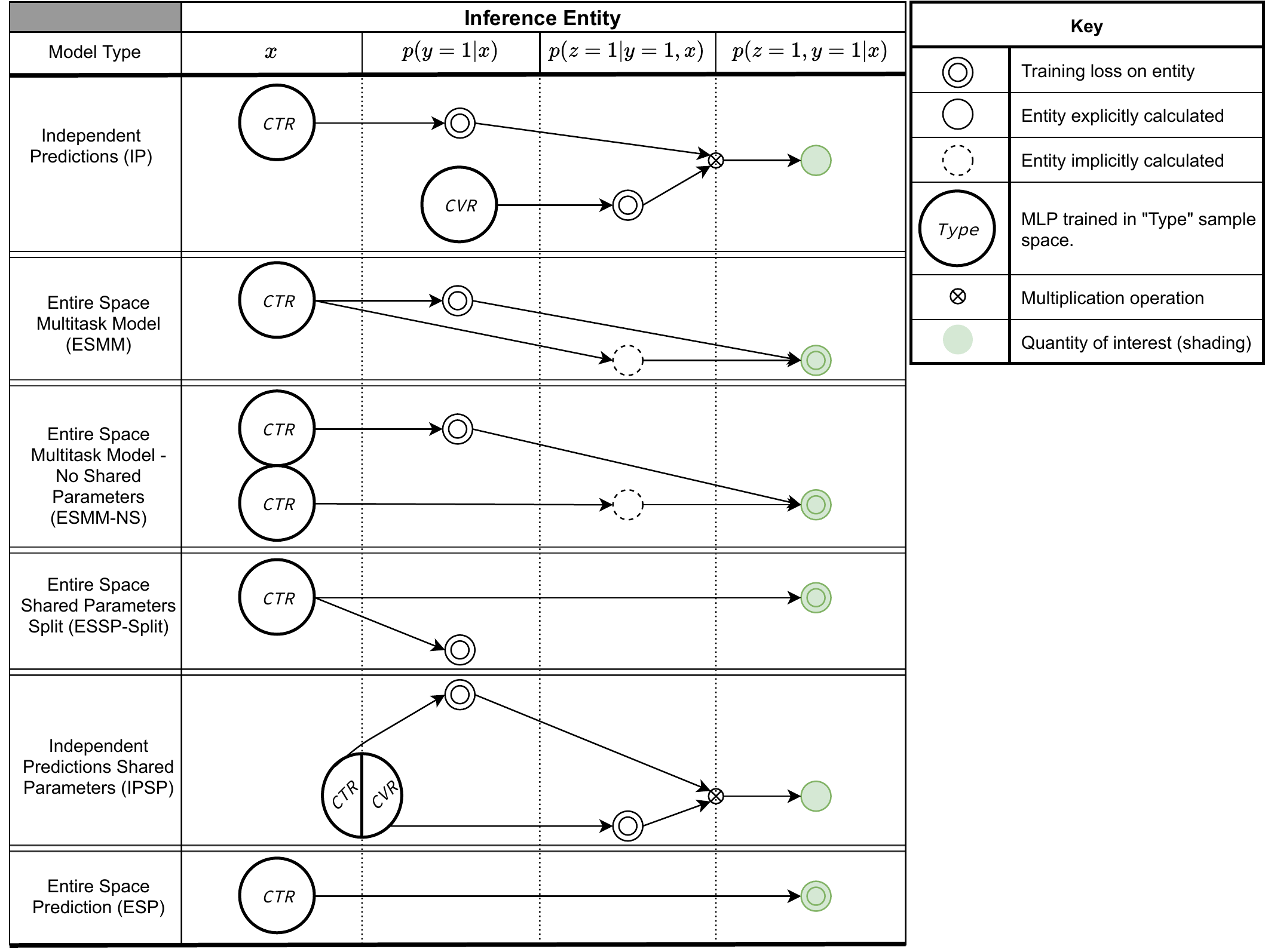}
    \caption{Different modeling approaches evaluated here. Independent Predictions (IP) serves as the baseline which models CVR and CTR as independent tasks. Entire Space Multitask Model (ESMM) includes all 3 characteristics, namely Shared Parameters, Entire Space prediction and Weighted CVR. The other models incorporate only some of these characteristics, allowing us to study their contributions to performance. The MLP labels (CTR, CVR) indicate the space of data used in training. CTR is all training examples (entire space), CVR is the subset of examples where $y=1$. (e.g. the ESMM-NS has two MLPs, both of which are trained on the CTR data). See Section \ref{sec:methods} for a description of each model.}
    \label{fig:models}
\end{figure}

In this work, we test 6 different approaches, including the na\"ive choice just described. Although MTL models have the potential to become complex we constrain our analysis to the use of (1) hard parameter sharing, (2) careful selection of training spaces and prediction heads, and (3) conditionally aware CVR prediction. We describe the 6 modeling approaches below\footnote{Table \ref{tab:models} in the Appendix provides a checklist-style summary of all our model designs, as well as some additional implementation details.} and also present this information for direct comparison in Figure~\ref{fig:models}.

We denote the baseline approach \textit{Independent Prediction} (IP) which treats CTR and CVR as separate tasks: two multi-layer perceptron (MLPs) with no shared parameters. CTR prediction, $\hat{p}(y=1|x)$, is trained on negative downsampled click data and CVR prediction, $\hat{p}(z=1|y=1, x)$, is trained using impressions that were clicked, $y=1$. The final prediction is constructed as the product of those two predictions, $\hat{p}(z=1,y=1|x) = \hat{p}(z=1|y=1,x) \cdot \hat{p}(y=1|x)$.

The primary approach introduced in \cite{ma2018entire} is to train a model to directly predict $\hat{p}(z=1, y=1|x)$ along with predicting $\hat{p}(y=1|x)$, and constructing the network such that $\hat{p}(z=1,y=1|x) = \hat{p}(z=1|y=1,x) \cdot \hat{p}(y=1|x)$. That is, there is an internal node in the network that can be considered as a prediction of $\hat{p}(z=1|y=1, x)$, but there is no loss directly optimizing this prediction. We refer to this approach as the \textit{Entire Space Multitask Model (ESMM)}, the name used by \cite{ma2018entire}. Our model is conceptually equivalent but the specific architecture is different (see Section \ref{sec:architecture}), in that we used hard parameter sharing in early DNN layers, as opposed to just the feature embeddings.

The ESMM approach introduces 3 characteristics distinct from the baseline (IP) approach:
\begin{itemize}
    \item ESMM uses hard parameter sharing between the CTR and CVR task. (Shared Parameters)
    \item ESMM trains the install prediction over the entire space of impressions by predicting $\hat{p}(z=1, y=1|x)$ rather than $\hat{p}(z=1|y=1,x)$. (Entire Space)
    \item ESMM implicitly weights the install prediction's loss by the click prediction. (Weighted CVR)
\end{itemize}

In order to separate the impact of these characteristics and understand their individual and combined effects we tested several variants of ESMM. \textit{Entire Space Multitask Model - No Shared} (ESMM-NS) uses the same losses as ESMM but has no shared parameters between the CVR and CTR prediction tasks. The \textit{ESSP-Split} model uses the same losses as the ESMM model and Shared Parameters, but the two predictions, $\hat{p}(y=1|x)$ and $\hat{p}(z=1,y=1|x)$, are made by independent heads with no constraint on their relationship\footnote{This means that the predictions of the two heads may be inconsistent since its possible for the model to predict $\hat{p}(z=1, y=1|x) > \hat{p}(y=1|x)$. In practice, this does not occur very often.}. \textit{Independent Predictions Shared Parameters} (IPSP) uses the same approach and losses as the IP model (that is, it is not an Entire Space model) but shares parameters between the CVR and CTR prediction. Finally, \textit{Entire Space Prediction (ESP)} just predicts $\hat{p}(z=1, y=1|x)$ with a single model, thus training over the whole space, but makes no use of the CTR task.

\subsection{Dataset and Training Setup}

The evaluation dataset for this paper is comprised of real click and conversion data for digital mobile app install ads served on Twitter, as well as MoPub, Twitter's mobile display network (e.g. in-game ads). While this real-world dataset allows us to evaluate the performance of these technologies on a truly representative problem the dataset itself is not publicly available due to numerous user privacy and business-sensitive constraints. Specifically, in each of the evaluations below a fixed dataset of click and conversion events collected during a consecutive number of days in mid 2020 were used for model training and evaluation. The raw data consisted of over 5 billion ad impressions (later down-sampled, as discussed below), over 50 million ad clicks, and several million conversion events\footnote{Specific dataset counts are approximated to avoid disclosure of proprietary information.}. Note, as discussed below, evaluation hold-out sets for these experiments always ensure past vs. future evaluation, such as training on the previous 14 days of data and testing on the 15th day. Also note, when training on the first $N$ days the examples are shuffled to make the data approximately i.i.d. 

Below, the results reported are for a single evaluation day. However, the robustness of these modeling approaches to temporal shift, i.e. how prediction performance changes as the model is tasked to make predictions further into the future without the benefit of retraining, were also evaluated. While this is a particularly interesting, and practically relevant, aspect of this problem we ultimately did not observe a noteworthy difference between the approaches in this regard.

\subsubsection{Negative downsampling}

Imbalanced datasets are a common problem in advertising datasets. We downsampled negative examples by some factor, $f$, and in order to calibrate the model, upweighted each negative sample by the same factor, $f$. Note that all samples where $y=1$ were kept, no downsampling was done based on the conversion label, $z$. The evaluation dataset was generated identically, with the same downsampling and upweighting procedure applied to negatives for the click task.

\subsection{Metrics}

Ultimately, for the purpose of ranking potential ads and valuing impressions, we are interested in the probability that an impression leads to a conversion, $p(z=1|x) = p(z=1, y=1|x)$. For this task we require predictions to be well-calibrated so we focused on the cross-entropy loss (we report PR-AUC in Appendix \ref{appendix:prauc}). We report our scores as relative percentage performance improvements versus the baseline model.

\subsection{Model architecture} \label{sec:architecture}

Since the models each have slightly different characteristics the exact architectures vary; however, the number of trainable parameters was kept comparable across all MLPs\footnote{IP and ESMM-NS have two MLPs and therefore about twice as many parameters overall.}. The multi-task models (except ESMM-NS) had two shared layers after the feature embeddings, followed by two layers \textit{per head} as the model branched. Models using a "Weighted CVR", e.g. ESMM, had the two branches reconnect with no trainable parameters after the $p(y=1|x)$ entity and the implicit $p(z=1|y=1, x)$ entity\footnote{The entity is "implicit" because there is no output for this value but the node can be interpreted as this prediction}, as in \cite{ma2018entire}. Models without this characteristic, e.g. IPSP, had a single entity at the root of each branch. We experimented with larger models, in terms of both wider layers and greater depth (more layers) for both the shared and branched parts of the network, but this did not bring any benefit. Larger models were also trained with batch normalization layers both included and excluded. The lack of benefit might be explained by lack of sufficient training data; though this is just another example of the wider open question about why larger models do not consistently perform better on recommendation tasks \cite{qin2021are}.

\section{Results}

We manually tuned all the models for similar numbers of experiments to find the best hyperparameters. In general, the models were fairly robust to hyperparameter choices, with the exception being the ESMM model which did have slightly more varied performance as a function of hyper-parameter values, discussed below.

\begin{figure}[b]
    \begin{minipage}{0.5\linewidth}
        \centering
        \includegraphics[width=0.7\textwidth]{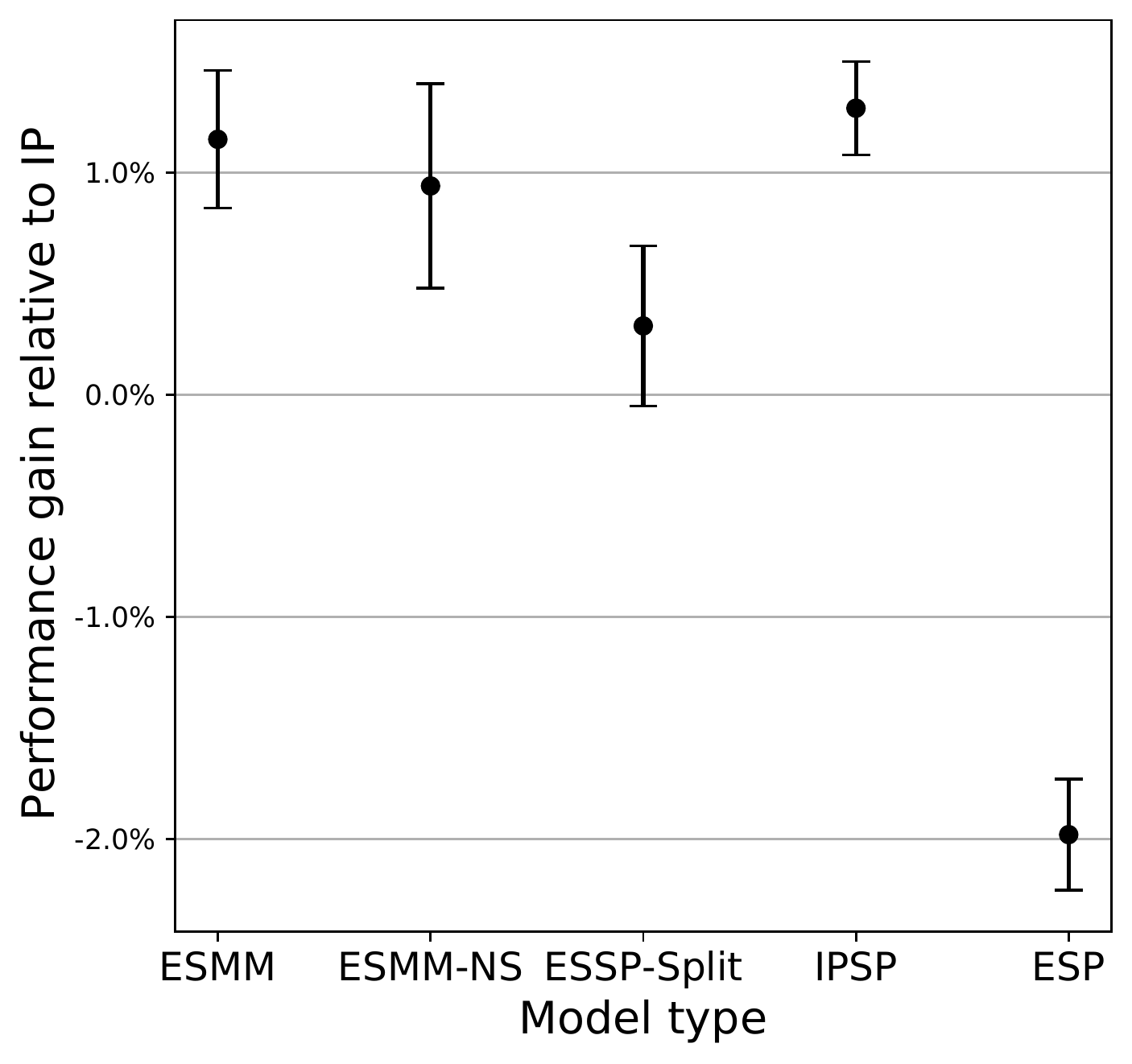}
    \end{minipage} 
    \begin{minipage}{0.4\linewidth}
        \begin{tabular}{lcl}
          \toprule
          Model & Performance $\pm$ SEM & Better Than \\
          \midrule
          IP          &  1 & ESP\\
          ESMM        & $1.0115 \pm 0.0628$ & IP, ESP\\
          ESMM-NS     & $1.0094 \pm 0.0932$ & ESP \\
          ESSP-Split  & $1.0031 \pm 0.0731$ & ESP \\
          IPSP        & $1.0129 \pm 0.0435$ & IP, ESP \\
          ESP         & $0.9802 \pm 0.0510$ & ESP \\
          \bottomrule
        \end{tabular}
    \end{minipage}
    \caption{Model performance by multi-task setup. Performance gains normalized by the mean score of the baseline IP model. The standard error for each model's performance difference (against IP mean) is calculated across at least 10 runs. The results show that there are multiple mechanisms for inducing positive transfer, but that hard parameter sharing alone (IPSP) may be optimal. Better than column indicates models that this model outperforms $p < 0.01$, 2-sided t-test.}
    \label{fig:results_plot}
\end{figure}

Figure \ref{fig:results_plot} gives a summary of the key results from our experiments. They provide clear evidence that a \emph{meaningful decomposition} of the prediction task has clear benefits, shown by ESP performing 2\% worse than IP. This na\"ive approach to training on the entire space of predictions leaves the model susceptible to learning noise when the positive install labels are so relatively infrequent, unaided by the useful signal that the click labels can provide.

There is then a performance jump to the ESSP-Split model, with a marked increase versus ESP. This comparison highlights the utility of hard parameter sharing. This is the `classical' benefit of MTL - which is often discussed in terms of "shared representations" or additional ``regularization'' \cite{ruder2017overview}. This same impact is demonstrated by IPSP, the best performing model, which kept the tasks as independent heads but leveraged combined early layer feature transformations. The benefit of the signal from the CTR task through shared parameters provided all of the gains in performance seen in alternative model designs.

Surprisingly, ESMM-NS performed very competitively. By simply weighting the loss on the CVR head by the click prediction, $\hat{p}(y=1|x)$, the model was able to perform better than baseline, and even beat ESSP-Split (not statistically significant). We suggest that what this highlights is the extent of the \emph{data bias} problem. If training on the entire space then there has to be some mechanism to assign `relevance' to the CVR samples -- otherwise, you get the poor performance of ESP. This, seemingly small detail, is (empirically) more important than any classical transfer learning arguments (since ESMM beats ESSP-Split). We do note that ESMM-NS increases the size of the model compared with all the other MTL designs, since, like baseline, it has two separate embeddings which is where most parameters exist even in very deep RecSys models. However, we don't suggest that the extra parameters are really helpful in this instance. In fact, given (over)fitting biased data seems to be an issue, more parameters alone would be likely to make things worse.

Finally, ESMM had similar performance to IPSP, albeit with greater variability. We tentatively suggest that with increased effort it may be possible to get consistent, larger performance gains from a well-tuned ESMM model. That is, we posit that the marginal benefit from increased model tuning for ESMM is much greater than for any of the other models. This would make sense given its design allows for the most complex learning interactions. But it is also a weakness in that simpler approaches may require less tuning.

An alternative view is that IPSP not training on the entire space may be positive since this avoids directly optimizing $p(z=1, y=1|x)$, which is arguably the most difficult training objective (i.e. having the worst signal-to-noise ratio).

\section{Conclusion}

We provide clear evidence that simple MTL methods can improve conversion model performance. Our experiments show that hard parameter sharing alone (IPSP) might be optimal for improving performance with significant, and relatively easy, wins versus a factored (IP) or na\"ive (ESP) baseline. We also establish the importance of counteracting the data bias problem that occurs when trying to predict installs on the entire space of impressions. The surprisingly simple solution of a weighted conditional install prediction tackles the bias well. However, we note that the gains from these two characteristics do not seem to be additive when combined. Whilst we studied this problem in the context of clicks and conversions, we suggest that this simple methodology can be used to explore other conditionally dependent tasks, as the methods to counter the fundamental problems of data sparsity and data bias should generalise well.

\section{Ethical considerations}

The research in the submitted paper has been reviewed as part of our organisation’s research and publishing process. This includes privacy and legal review to help ensure that all necessary obligations are satisfied.

As with many companies that rely on advertising to fund free and open access to products and services, our platform utilizes algorithms that recommend personalized content, including ads.  Recommender systems are imperfect, and automated decision systems may not treat all people equitably.  The identification and prevention of inequity and bias in ML is a growing field of research that we closely follow.  

Despite ongoing efforts to detect and prevent algorithmic amplification of bias, inequality still exists in society and therefore may impact the source data used to train many models. The authors of this paper are not aware that the experiments conducted resulted in any positive or negative impacts on the inherent bias that exists in recommender systems.


\bibliographystyle{ACM-Reference-Format}
\bibliography{multitask}

\appendix
\newpage

\section{Model characteristics checklist}

\label{appendix:characteristics}

Table \ref{tab:models} provides a summary of the design choices made for each model. We also provide some additional description below to remove any ambiguity surrounding how the models were implemented.

\begin{table}[h!]
    \centering
    \begin{tabular}{c|c|c|c}
         Model Name & Shared Parameters & Entire Space & Weighted CVR \\
         \hline
         IP & & & \\
         ESMM & \checkmark & \checkmark & \checkmark \\
         ESMM-NS &            & \checkmark & \checkmark \\
         ESSP-Split & \checkmark & \checkmark &     \\
         IPSP & \checkmark &            &             \\
         ESP &      & \checkmark & \\
    \end{tabular}
    \caption{The 6 different models tested and the ``characteristics'' which each model captures.}
    \label{tab:models}
\end{table}

The IP model, our baseline, uses two completely disjoint MLPs. One model is trained with a dataset of impressions, $x$, and predicts clicks, $y$. The other model is given a dataset of clicked impressions, $y=1$, and predicts installs, $z$.

For the rest of the models a single dataset containing \emph{downsampled impressions and all clicks} was used. We then use sample weights on the respective loss heads to produce the training regime required, along with model design, discussed in Section \ref{sec:methods}.

ESMM, ESMM-NS, and ESSP-Split use all the training samples on both heads. That is, the sample weight is set to $1.0$ for every sample for both losses.

The IPSP model requires setting some sample weights to $0.0$. Specifically, any impression that was not clicked, $y=0$, had a sample weight of zero for the CVR prediction, $p(z=1|y=1,x)$. As a consequence, only the parameters in the CTR branch and the shared parameters (via the CTR branch) would be updated for these unclicked samples. This does mean that for a set batch size, the number of samples generating gradient updates via the CVR loss is (1) variable and (2) smaller than the batch size\footnote{(2) holds in all cases, other than the vanishingly low probability event that all the samples in a batch were clicked.}.

The ESP model uses a single dataset, requiring only the install label, $z$, with all sample weights set to $1.0$.

\section{PR-AUC} \label{appendix:prauc}

\begin{figure}[h!]
  \centering
  \includegraphics[width=0.5\textwidth]{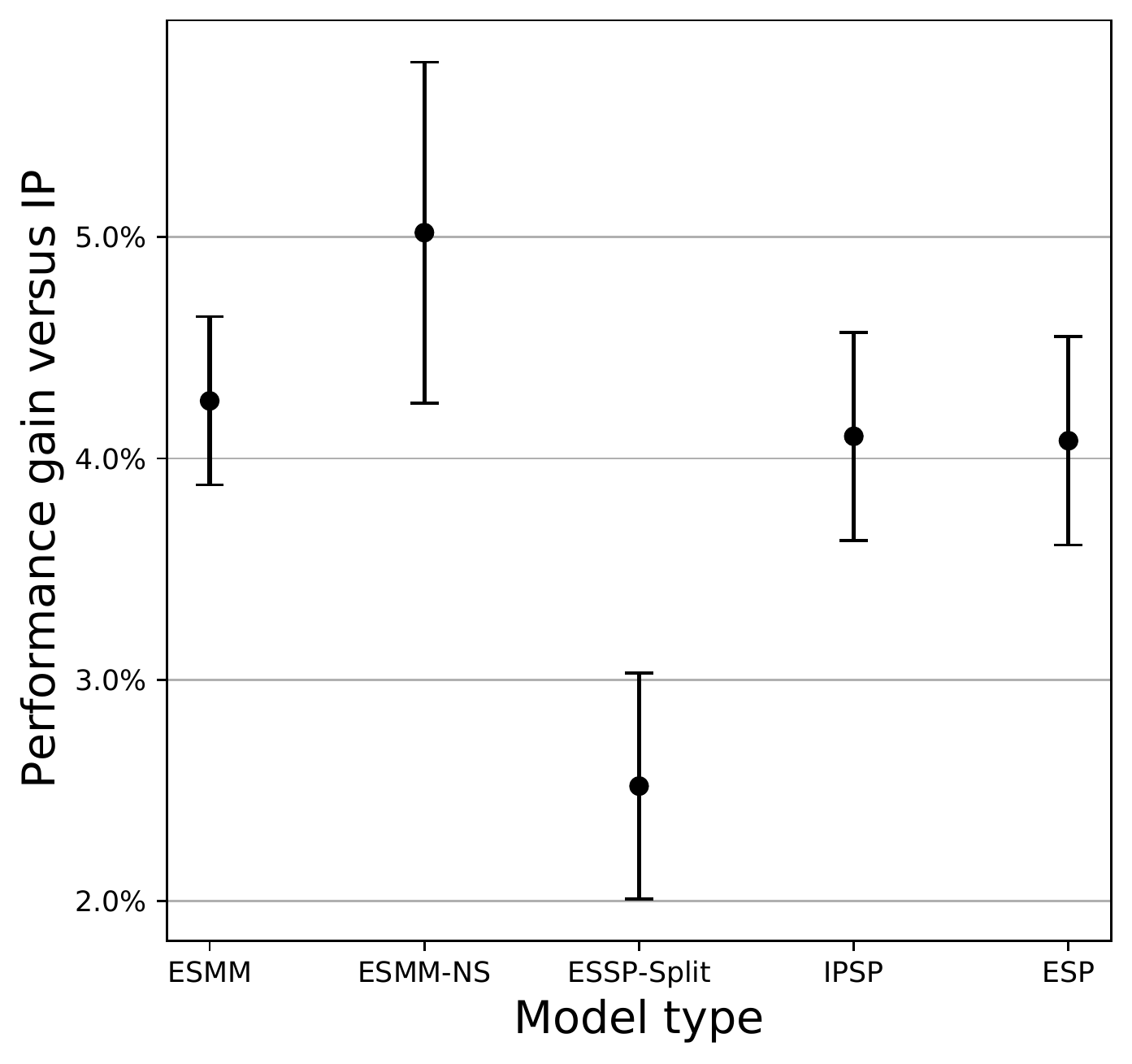}
  \caption{PR-AUC performance of models.}
  \label{fig:prauc}
\end{figure}

We focused on the cross-entropy metric, because for many online advertising applications the calibration of the model is important. However, for completeness we include the results with PR-AUC metric in Figure \ref{fig:prauc}. Although the ordering of the mean shifts compared with the CE metric, the overall conclusion is unchanged. Several different approaches outperform the baseline IP approach, and all shared parameter or entire space models perform fairly well. Notably, ESP performs much better on this metric.

\section{Non-stationarity}
\label{appendix:non-stat}

We thought it might also be interesting to observe the performance degradation of the model’s predictions. In general our evaluation metrics were calculated on the next day of data, i.e (training+1)-th day. For these experiments we wanted to observe what happened as we increased the period of time between training and predictions. The idea motivating the experiment was that there may be some difference in the way the models learn (e.g. one possibility being that the MTL model is forced to learn better user embeddings that, \emph{possibly}, could generalise better over time). The comparison here was between the IP and ESMM model designs.

\begin{figure}
    \centering
    \includegraphics[width=0.9\textwidth]{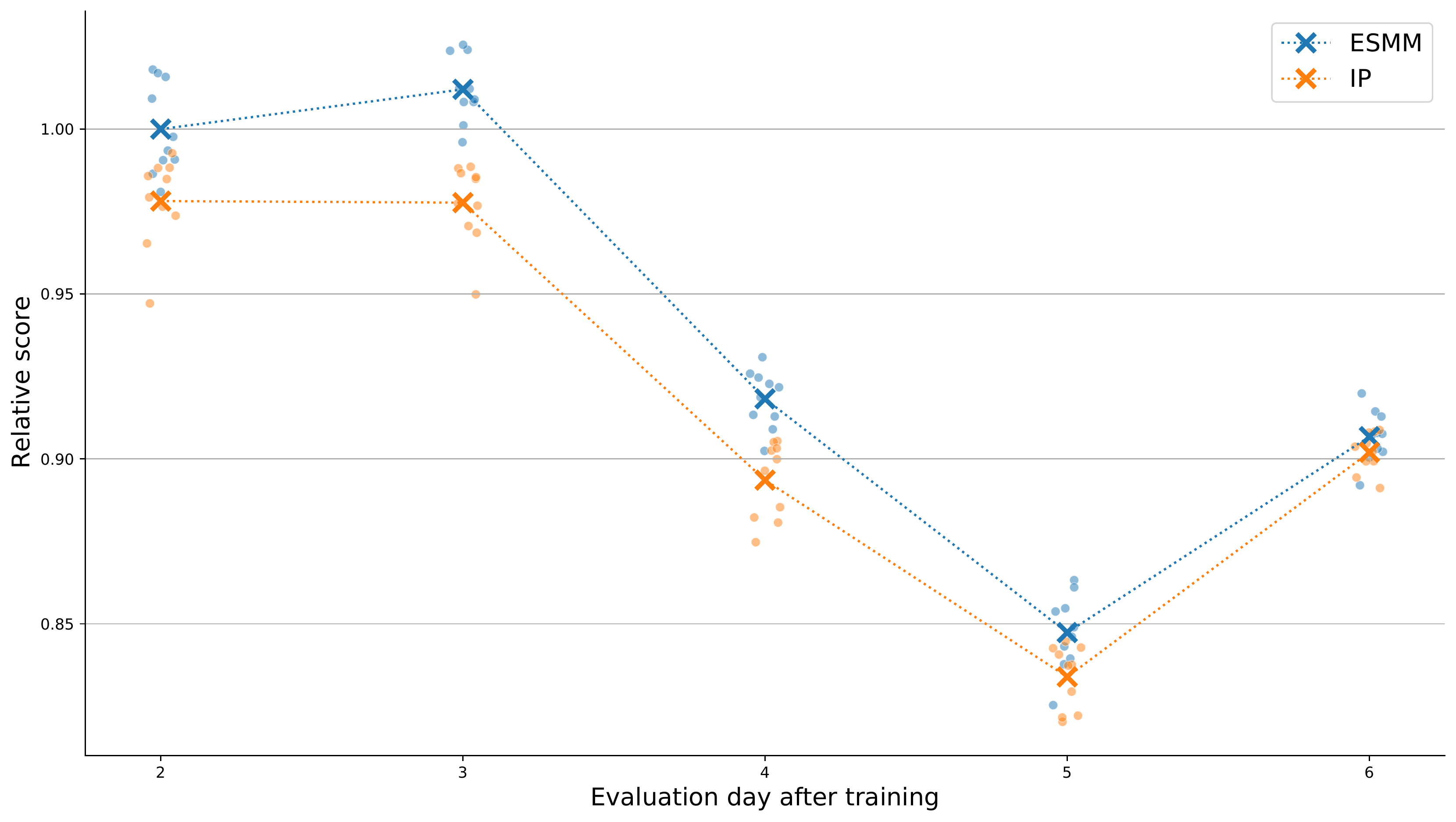}
    \caption{Model performance decay}
    \label{fig:model_staleness}
\end{figure}

The data in Figure \ref{fig:model_staleness} was generated by training 10 models of both types (each with their own set of tuned hyperparameters) and evaluating each of these models on the $n$-th day of data after training for $n=2, 3, ..., 6$. The cloud of points are the scores for each model on a given day and the X marker denotes the mean (the variance is not significantly different so we omit it from this plot). We did not observe any meaningful patterns or behaviour of the performance delta. The ESMM model does retain its performance win over the IP model, albeit by Day 6 this is practically zero, but the prediction performance of both models seems to decay similarly. Note, there is significant inter-day variation in the performance which explains the gains seen between day 5 and 6, and the small (average) improvement for ESMM between day 2 and 3. These ``improvements'' are just fluctuations in the data and not a consequence of any modelling decisions: put differently, such patterns would likely emerge (stochastically) with any type of classifier.

\section{CTR task}
We have assumed throughout that installs, $p(z=1|x)$, are the quantity of interest, and that clicks, $p(y=1|x)$, are of no interest, except insofar as they are relevant for the install prediction. We note that CTR performance in these models fluctuates dramatically. ESP, for example, does not even predict CTR, and any model featuring Weighted CVR performs badly for CTR. This should be no surprise as the gradient from the CVR head is partially backpropagated through the CTR branch via the multiplication operation. We note this because engineers, or teams, that have a business or machine learning motivation to accurately predict CTR will (1) either have to train a separate model specifically for this task or (2) accept the performance penalty for certain model designs.

\end{document}